# Why Do We Believe in the Second Law?

Todd L. Duncan

*Science Integration Institute, 1971 SE 73rd Ave., Hillsboro, Oregon, 97123, and Portland State University Center for Science Education, Portland, Oregon 97207*

**Abstract.** Claims of exceptions to the second law of thermodynamics are generally met with extreme skepticism that is quite reasonable given the great confidence placed in the second law. But what specifically is the basis for that confidence? The perspective from which we approach experimental or theoretical results that call into question the absolute status of the second law depends greatly on our understanding of why it must be true. For example, a belief that there are solid theoretical arguments demonstrating that the second law *must* be true leads to a very different perspective than a belief that the law is simply a generalization of empirical observations. This paper will briefly survey and examine some of the basic arguments on which our confidence in the second law might be based, to help provide a well-informed perspective for evaluating the various claims presented at this conference.

## INTRODUCTION

The second law of thermodynamics is one of the most fundamental and well-established principles of physics. Although originally formulated as simply a generalization of observational experience about the way in which heat flows and our inability to construct perpetual motion machines of the second kind, it has attained a status and domain of application that extends far beyond its direct observational foundations. This status is conveyed in Eddington's well-known statement: "The law that entropy always increases—the second law of thermodynamics—holds, I think, the supreme position among the laws of Nature." [1]

Most physicists today share a similar confidence in the universal validity of the second law, supported by its successful application to a wide variety of situations well outside the realm of heat engines in which it was developed. But it is striking that there is no such widespread agreement about the fundamental *justification* on which this confidence is based, nor even about the exact interpretation of the precise restriction the second law imposes upon processes in nature. Despite over 130 years of attempts to justify or "prove" the second law of thermodynamics in some form, no single argument has emerged as clearly satisfactory. Arguments vary widely from person to person and often depend on subtle or hidden assumptions and preconceptions, and it can be difficult to sort out what has actually been proven and the exact conditions under which a particular justification of the second law applies. This situation has taken on a greater urgency with the recent string of experimental and theoretical challenges to the second law (see *e.g.* [2-4]). In order to fairly evaluate these sorts of challenges, one needs a clear understanding of the foundation on which one's belief in the second law rests.

An important benefit of a meeting such as this is the opportunity to rethink some of our assumptions and prejudices from as broad a perspective as possible. Our faith in the second law is most likely justified, but the grounds for our confidence are often poorly articulated. This paper presents the framework of my own efforts to understand where this confidence comes from and what assumptions it depends upon. Most of the information given is probably familiar, but the presentation may suggest new insights or at least new perspectives, because here it is organized around the specific question, "Why do we believe in the second law?"

Arguments for why it must be true obviously depend on what the law actually says. It's convenient to divide statements of the second law, and the corresponding arguments for its validity, into 3 categories: 1) forms of the law which are essentially formalizations of direct empirical observations; 2) forms based on statistical treatments of the underlying mechanics as the foundation for the second law; 3) forms whose basis rests on information theory.

## FORMALIZATIONS OF DIRECT EXPERIENCE

One perspective is to view the second law as simply a formalized generalization of the very common and concrete observation that heat flows spontaneously from hot bodies to cold ones, and will not spontaneously flow in the other direction. The formalization of this general tendency came about during the nineteenth century through the efforts to understand the limitations on the performance of heat engines [5]. From this perspective, the argument for the universal validity of the second law rests on the sheer weight of observational experience about the limitations on obtaining mechanical motion from heat, combined with aesthetic arguments about the simplicity and beauty of the thermodynamic theory that unifies these observations.

Carnot provided the initial formalization of these observations by focusing on the key issue: "The question has often been raised whether the motive power of heat is unbounded, whether the possible improvements in steam-engines have an assignable limit—a limit which the nature of things will not allow to be passed by any means whatever; or whether, on the contrary, these improvements may be carried on indefinitely." [6] While previous attempts at addressing this question had been specific to particular machines, Carnot's insight was "to consider in the most general way the principle of production of motion by heat…independent of any mechanism or any particular agent. It is necessary to establish principles applicable not only to steam-engines, but to imaginable heat-engines, whatever the working substance and whatever the method by which it is operated." [6]

Thomson's (Kelvin's) formulation retains this flavor of a limitation on what useful work can be achieved by the transfer of heat. His statement of the second law reads: "It is impossible, by means of inanimate material agency, to derive mechanical effect from any portion of matter by cooling it below the temperature of the coldest of the surrounding objects." [7]

In 1865 Clausius [8] completed the formalization of the second law into precise mathematical form by introducing the experimental or thermodynamic entropy $S_e$, defined by $dS_e = dQ/T$ (where $dQ$ is the increment of heat transferred from a heat bath at temperature $T$ as defined by Kelvin's scale) and stating the second law as the now-familiar restriction that processes in nature occur in such a way that the "entropy of the universe tends toward a maximum." This form is particularly convenient for our discussion because any version of the second law can be expressed as a statement that the total entropy of the universe is a nondecreasing function of time, allowing us to focus attention on the different ways of thinking about the underlying meaning of entropy. In a sense we can see the progression of categories of the second law discussed in this paper as providing progressively deeper understanding of what entropy means. Clausius' definition in terms of heat flow *(dQ/T)* reflects the direct link to observation. At this stage, there was no deeper understanding of what the entropy represented. (See *e.g.* [9] for a more precise and complete discussion of Clausius' formulation of the second law).

From this "empirical formalization" perspective, one's confidence in the validity of the second law ultimately rests on the observation that perpetual motion machines are impossible or, equivalently, that heat will not flow spontaneously from a cold body to a hotter one. The formalization of these observations provided the beautiful, simple, and powerful theoretical construct of thermodynamics which is tremendously successful in predicting a great variety of experimental results relevant to everyday experience. Hence it seems very unlikely that any of its fundamental principles such as the second law can be incorrect within this realm of experience. In this sense it is similar to any of the classical laws of physics such as Maxwell's laws of electromagnetism or Newton's law of gravitation. Of course, if the domain of application is stretched far enough, Newton's law of gravitation, for example, no longer works and must be replaced by general relativity. Similarly, we must admit from this perspective the possibility that the second law may break down and be replaced by a more comprehensive law under certain extreme conditions. So in evaluating challenges to the second law from this perspective, we might focus our attention on whether there are extreme conditions involved (perhaps including, for example, conditions where quantum coherence effects are important) that may justify a closer look and an open mind to the possibility that the challenge might be valid.

# FORMS BASED ON STATISTICAL DYNAMICS

A second general perspective one may take is that the second law follows in some natural way from a more fundamental microscopic theory. This would be more satisfying and more convincing since from the perspective of the last section, one naturally wonders *why* heat flows from hot to cold and not vice versa, or why Clausius' mysterious entropy never decreases with time.

Boltzmann laid the groundwork for such an underlying foundation by applying statistical arguments to the then still controversial kinetic/molecular theory of gases (see [5] for a thorough discussion of this history). Since the huge number of molecules involved in typical macroscopic systems make it impossible to follow the dynamics in detail, he made a series of reasonable statistical assumptions to derive an equation for the time evolution of the distribution function *f(x,p,t)* representing the number of particles with various momenta at various locations in the system. Boltzmann then cleverly defined a quantity now referred to as $H \propto \int (f \ln f) dx dp$, which is related to the phase space volume corresponding to a given *f* for the ideal gas systems he considered. He then used his statistical equation for the time evolution of *f* to show that *H* can only decrease with time, thus providing a suggestive argument for how the second law could follow naturally from the microscopic behavior of a large number of particles. The argument is only suggestive because as Jaynes [10] points out, Boltzmann's *H* is not directly related to Clausius' thermodynamic entropy $S_e$, and there are approximations made in the process of substituting statistical approximations for the "true" microscopic dynamics of the system which confuse the issue of the extent to which one has "derived" a second law - like expression for *H* without essentially putting in the irreversibility by hand. For various perspectives on the status of Boltzmann's *H*-theorem and more modern versions of this form of justification for the basis of the second law, see *e.g.* [11-13].

More generally, this approach provides a way to see the second law as a natural outcome of the reasonable expectation that systems evolve through macroscopic (observable) states occupying progressively larger volumes in phase space. This is expressed in Boltzmann's more famous general identification of entropy with phase space volume, $S_B = k \log W$, where *W* is the volume in classical phase space corresponding to a particular observable macroscopic state of a system, and *k* is the Boltzmann constant. From this perspective the second law, which seemed so mysterious from the perspective of the last section, can seem so obvious as to be almost trivial. Of *course* systems will evolve in such a way that they will probably be observed in their most probable macroscopic states. In addition, if the probabilities are sharply peaked as they are in the limit of very large numbers of particles, then the high probability of observing the system in particular states becomes a near certainty.

Solidifying this argument involves justifying that the actual dynamics of the system (the orders that nature is *actually* following to produce the time evolution of the real system) are in fact well enough approximated by the statistical assumptions in our models. The question has been shifted from "Why does heat flow from hot to cold?" to "Why does a probability argument work for the real-time dynamics of the system?" For a discussion of the status of the justification for this so-called ergodicity of dynamical systems see *e.g.* [14]. From this perspective challenges to the second law might be seen in terms of whether there exist systems in which the dynamics does not produce an evolution of the system through macrostates corresponding to steadily increasing phase space volume (or number of microstates in the quantum version).

## FORMS BASED ON INFORMATION THEORY

Szilard's 1929 analysis [15] of the Maxwell's Demon puzzle paved the way for a surprising connection between thermodynamics and the concept of *information*, which leads to a third class of second law formulations pioneered by Brillouin, Jaynes, and others. By identifying entropy with "missing information," the second law can be stated as a requirement that the total amount of missing information in the universe be a nondecreasing function of time.

Of course, we need to know what "missing information" means and why this is related to the thermodynamic entropy $S_e$ before we can make much use of this connection. Articles by Machta [16] and by Jaynes [17,18] are recommended for an introduction to the topic. Qualitatively the amount of information describes the extent to which one has well-defined knowledge of the state of the system (or equivalently, answers one has to questions about the system). Information is measured in terms of the probability, $p_i$, of a given state compared to all possible states that are consistent with the constraints on a system. While there are many ways to formally define an information-based entropy, Shannon's [19] $S_I = -\sum_i p_i \log p_i$ provides a clear connection to the Gibbs entropy commonly used in thermodynamics.

The important point for the present purpose is that once a connection between $S_e$ and $S_I$ is established, we have a much deeper way of approaching the central question of this paper. Now the question, "Why does the second law hold?" can be translated to, "Why does missing information increase with time?" and this shifts our thinking significantly. Rather than puzzling over why heat flows as it does, we're now puzzling over how nature defines questions that can be asked, the number of answers that can be given to those questions, the extent to which the answers are known in some way, and how these things change with time. So in analyzing situations that may challenge the second law, we have a new way to look at them, in terms of whether there is some aspect of the system that becomes better defined without a compensating loss of information somewhere else. This opens up new avenues for understanding these systems and for seeing the fundamental restriction expressed through the second law.

## CONCLUSION

If the second law is stated in the form, "the entropy of the universe is a nondecreasing function of time," then one's belief in the universal validity of the law depends upon an interpretation of what entropy is, and the corresponding argument for why that entropy should not decrease. The second law challenges presented at this conference and elsewhere in the recent literature provide an opportunity for each of us to better articulate our own interpretations and arguments and to clarify why our confidence in the second law is well founded. Whatever perspective we take should include a clear idea of what new evidence would present a legitimate challenge to the second law as we have articulated it for ourselves, or else we risk falling into the trap of believing in the law as a matter of faith, without understanding why.